# High fill factor confocal compound eyes fabricated by direct laser writing for better imaging quality

Haodong Zhu,[1] Junyu Xia,[1] Yi Huang,[1] Minglong Li,[1] Guangqiang He,[2] Ming Zhao,[1*] and Zhenyu Yang[1]

[1]*Nanophotonics Laboratory, School of Optical and Electronic Information, Huazhong University of Science and Technology, Wuhan 430074, China*
[2]*State Key Laboratory of Advanced Optical Communication Systems and Networks, Department of Electronic Engineering, Shanghai Jiao Tong University, Shanghai 200240, China*
*Corresponding author: zhaoming@hust.edu.cn*

**We fabricate two kinds of 100% fill factor compound eye structures using direct laser writing, including conventional compound eyes (CVCEs) with the same focal length of each microlens unit, and specially designed confocal compound eyes (CFCEs). For CFCEs, the focal length of each microlens unit is determined by its position and is equal to the distance between the microlens unit and the image sensor. In this letter, the optical properties of CVCEs and CFCEs are tested and compared. It is found that compared with CVCEs, CFCEs can improve the focusing efficiency by about 7%, enlarge the imaging area by about 25%, and have better imaging quality at the edge of the field of view.** © 2021 Optical Society of America

After billions of years of evolution, creatures in nature have evolved a variety of interesting structures and functions. In recent decades, many scientists have obtained a series of excellent results from nature [1-4]. Among these, the artificial compound eye structure with the wide field of view (FOV) and low aberration, which imitates natural compound eye vision system, has attracted extensive attention and become an important research topic in the field of optical bionics [5-8].

The natural compound eye of insects consists of an external imaging system and an internal sensing system. The imaging units (ca. 10–30 μm) in different areas are imaged on their respective photoreceptor cells which form a complex light-sensing curved surface. Therefore, for artificial compound eye, the microlens array imitating the external imaging system of natural compound eye needs to be designed on a curved substrate. However, mature image sensors are generally planar structure due to the limitations of fabrication technology and materials, which causes a difference in the distance between each microlens and the plane of the image sensor. In the conventional artificial compound eye structure, each microlens unit has the same parameter, so it is difficult for each microlens to image clearly on the target surface of the image sensor at the same time. To solve this problem, some researchers have added an optical relay system between the artificial compound eye structure and the image sensor to convert the curved focal plane of the curved microlens array into a planar focal plane [9]. However, this method requires multiple optical elements, which not only has calibration problems but also increases the volume of the system.

There are many methods to fabricate compound eye, such as diamond turning [10-11], surface modification [12], hybrid sol-gel method [13], and thermomechanical process [14]. However, the first method cannot be used to fabricate micron-size microlens units. The other three methods are difficult to achieve precise control of the compound eye structure. two-photon photopolymerization direct laser writing (DLW) [15-18] is considered by researchers to be an effective method for fabricating complex three-dimensional (3D) micro-nanostructures due to its excellent processing accuracy that break the classical optical diffraction limit and 3D rapid prototyping capacity, and has a wide range of applications in micro-optics [19,20], micromechanics [21], microfluidics [22] and so on.

In this letter, we choose DLW to fabricate two artificial compound eye structures, including conventional compound eyes (CVCEs) with the same focal length of each microlens unit, and specially designed confocal compound eyes (CFCEs), where the focal length of each microlens unit is determined by its position and is equal to the distance between the microlens unit and the image sensor. Both structures have a 100% fill factor, and the size of the microlens unit is about 10.5 μm. The research shows that compared with CVCEs, CFCEs can improve the relative focusing efficiency by about 7%, enlarge the imaging area by about 25%, and have better imaging quality at the edge of FOV. Moreover, since the CFCEs adopts a single-layer structure, there is no need to add optical relay system, which makes the whole system simple, compact and stable.

Two structures of artificial compound eyes were designed, which are both composed of hexagonal microlens arrays arranged on a spherical substrate with a 100% filling factor. The spherical substrate height is $H = 20$ μm, the radius is $R = 50$ μm, and the size of the microlens unit is about 10.5 μm, as shown in Fig. 1. The differences of two artificial compound eyes are: the focal length ($f = 24$ μm in this letter) of each microlens unit in the compound eye structure shown in Fig. 1(a) is homogeneous, and their focuses are on the curved surface $S_2$, which is the same as the common compound eye structure. Therefore, we name it conventional compound eyes, and its model is shown in Fig. 1(c). Note that the dimensions are the size of the overall structure, including the spherical substrate and the protruding microlens array. In Fig. 1(b), the focal length $f_n$ of each microlens unit is determined by its position and is equal to the distance $l_n$ between the microlens unit and the image sensor, that is $f_n = l_n$. Since the substrate is spherical, according to the geometric relationship, one can get the $l_n$ as follows:

$$l_n = R - \frac{R - l_0}{\cos(\theta_n)}, \quad (1)$$

where $R$ is the spherical radius, $l_0$ is the distance between the central microlens unit and the image sensor, and $\theta_n$ is the angle between the central axis of the compound eye and the optical axis of any microlens unit. Combined with the single spherical imaging formula, the curvature radius of the microlens unit can be obtained as follows:

$$\rho_n = (R - \frac{R - l_0}{\cos(\theta_n)})(1 - \frac{1}{n_0}), \quad (2)$$

where $n_0$ is the refractive index of the material. In order to ensure consistency with CVCEs, we design $f_n = l_n = 24$ μm. The focus of each microlens unit of this specially designed compound eye is on the planar plane $S_1$. Therefore, we name it conventional compound eyes, and its model is shown in Fig. 1(d).

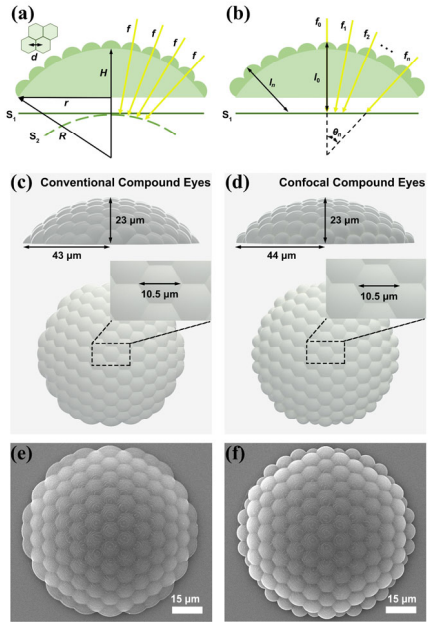

**Fig. 1.** Artificial compound eyes fabricated by DLW. (a) and (b) The schematics of CVCEs (left) and CFCEs (right), respectively. (c) and (d) The models of CVCEs (left) and CFCEs (right), respectively. (e) and (f) Top-view SEM images of CVCEs (left) and CFCEs (right), respectively. The scale bar in the image is 15 μm.

According to the above design, two kinds of compound eye structures are fabricated by our DLW system. A 100× oil-immersion objective lens (Nikon) is used to focus the laser beam with a central wavelength of 780 nm, a pulse width of 100 fs and a repetition rate of 100 MHz (C-Fiber 780, Menlo Systems GmbH) into photoresist IP-S (Nanoscribe GmbH). The system uses galvanometer scanner (Intelliscan III 10, Scanlab GmbH) and piezoelectric displacement platform (P-563.3CD, Physik Instrumente) to realize fast laser 3D scanning. To attain high resolution and processing efficiency, a writing speed of 100 mm/s and an average laser power of 27 mW are found to be optimal after many repeated attempts. When the scanning is finished, the structures are immersed in the propylene-glycol-monomethyl-ether-acetate for 30 min to remove the unpolymerized photoresist.

After fabrication, the structures of CVCEs and CFCEs under a scanning electron microscope are shown in Figs. 1(e) and 1(f), respectively. The surface of CVCEs and CFCEs are smooth, and the size of them are 86.4 μm and 88.5 μm, respectively, which deviates from the design value of 86 μm and 88 μm by 0.47% and 0.57%.

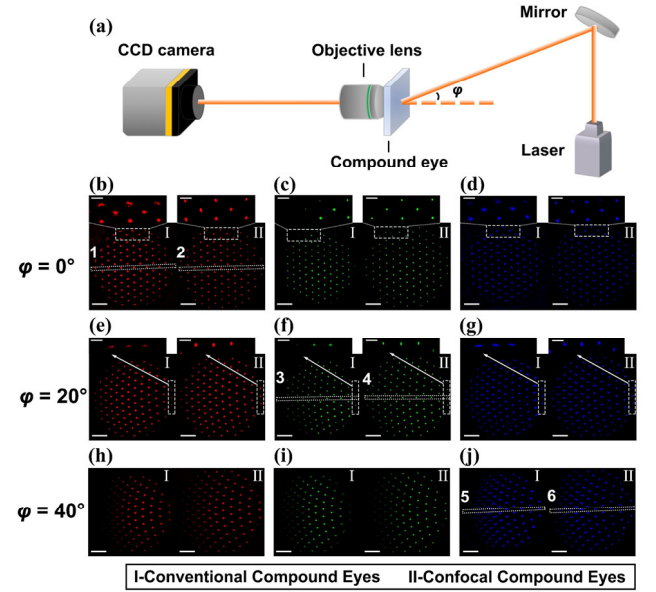

**Fig. 2.** (a) Experimental setup for optical focusing. (b)-(j) Optical focusing properties of CVCEs and CFCEs. The left side of each subimage is CVCEs and the right side is CFCEs. The incidence angles from top to bottom are $\varphi = 0°$ ((b)-(d)), $\varphi = 20°$ ((e)-(g)) and $\varphi = 40°$ ((h)-(j)), respectively. The central wavelengths of the laser are 650 nm (red), 532 nm (green) and 473 nm (blue), respectively. The scale bar in the image is 15 μm, and the scale bar in the inset ((b)-(g)) is 5 μm.

In order to character the optical properties of the compound eye structures, an optical system composed of three kinds of semiconductor lasers, a 100× objective lens and a charge coupled device (CCD) camera is set up, as shown in Fig. 2(a). By adjusting the mirror, the collimating laser beams with wavelengths of 473 nm, 532 nm or 650 nm incident into the two kinds of compound eye structures at different incident angles, and a CCD camera is installed to capture the image magnified through the objective lens. The results are shown in Figs. 2(b)-(j), where the left side of each subimage is CVCEs and the right side is CFCEs respectively. Figs. 2(b)-2(d) show the focal spots array when the beam is normal incident ($\varphi = 0°$, $\varphi$ is the angle between the central axis of compound eye and incident light). It can be found that the two kinds of compound eye structures have good optical focusing ability. As can be seen from the inset in Figs. 2(b)-2(d), CVCEs cannot focus effectively at the edge of FOV, while the focal spots of CFCEs still keep clear and sharp. Figs. 2(e)-2(j) show $\varphi = 20°$ and $40°$ light beams received by CVCEs and CFCEs. In the case of large incident angle, the compound eye structure still has some areas that can focus without significant distortion, which is different from the single lens. In addition, with the increase of incident angle, the effective focusing area and the number of clear focal spots will gradually decrease. It can be seen that CFCEs is better than CVCEs in both effective focusing area and focusing quality at the edge of FOV.

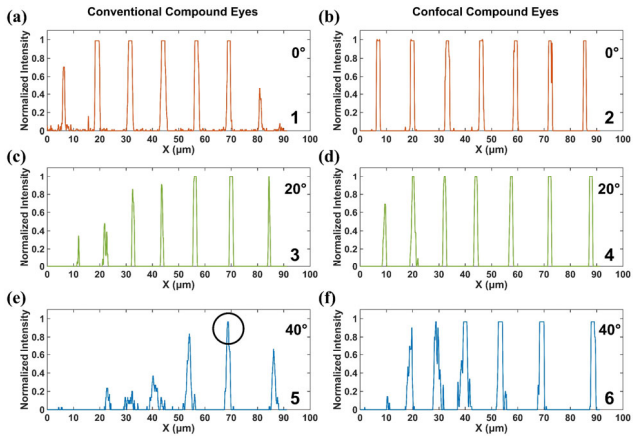

**Fig. 3.** Grayscale intensities of the six groups of focal spots from Figs. 2(b), 2(f) and 2(j). On the left is CVCEs, and on the right is CFCEs. The incidence angles and the central wavelengths of the laser from top to bottom are $\varphi = 0°$ and 650 nm (red) ((a) and (b)); $\varphi = 20°$ and 532 nm (green) ((c) and (d)); $\varphi = 40°$ and 473 nm (blue) ((e) and (f)), respectively.

To quantify the optical focusing performance of CVCEs and CFCEs, six groups of focal spots were selected from Figs. 2(b), 2(f) and 2(j) to obtain the normalized grayscale intensities, as shown in Fig. 3. In this case, note that there are seven microlens units (from left to right are $\theta_n = 52.5°, 35°, 17.5°, 0°, 17.5°, 35°, 52.5°$) to focus the light beam and ideally there should be seven focal spots in focal plane. It can be seen from Figs. 3(a) and 3(b) that when the incident angle is $\varphi = 0°$, the outermost microlens unit ($\theta_n = 52.5°$) of CVCEs cannot focus effectively, while the focal spots of CFCEs are still sharp and the peak intensity is barely weakened. As shown in Figs. 3(c)-3(f), with the increase of incident angle, the focusing quality of CVCEs and CFCEs decreases obviously, and the microlens unit with smaller angle between its own optical axis and incident light has better focusing effect. For example, the angle between the optical axis of the microlens unit corresponding to the highest peak (black circle) in Fig. 3(e) and the incident light is 5°. Moreover, it is found that the focusing quality of CFCEs is significantly better than that of CVCEs, and CFCEs has more complete focal spots. This is because when used at oblique angles of incidence, the optimal CCD imaging plane of CVCEs needs to be moved down to match the focal length of the microlens unit with a small angle between optical axis and incident light, which will significantly affect the focusing quality of the microlens unit with a large angle. However, for CFCEs, confocal design makes the focus of each microlens unit be in a planar plane, which makes up for above defect to some extent and achieves better focusing quality.

Furthermore, to analyze the overall focusing characteristics of the two compound eye structures, this letter calculates the relative focusing efficiency [23]:

$$\eta = \frac{I_{\text{Sp}}}{I_{\text{All}}}, \quad (3)$$

where $I_{\text{Sp}}$ is the focal spot energy, $I_{\text{All}}$ is the total transmitted energy. The Fresnel reflection loss and absorption loss are ignored in this calculation. Under different incident light, the measured focusing efficiencies of CVCEs are 84.7% ($\varphi = 0°$), 67.7% ($\varphi = 20°$), 49.6% ($\varphi = 40°$), and these of CFCEs are 88.5% ($\varphi = 0°$), 74.8% ($\varphi = 20°$), 58.4% ($\varphi = 40°$). It is obvious that due to the hexagonal 100% fill factor design, both CVCEs and CFCEs have high focusing efficiency. What's more, as the incidence angle increases, the focusing efficiency of compound eyes decreased significantly, and the focusing efficiency of CFCEs is higher than that of CVCEs about 5%-9% under the same conditions.

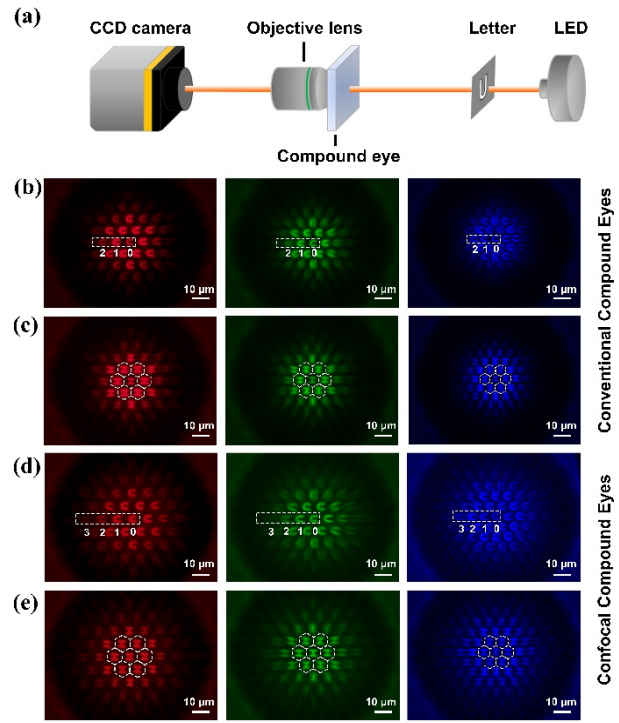

**Fig. 4.** (a) Experimental setup for optical imaging. (b)-(e) Imaging of the letters "U" and "H" using CVCEs and CFCEs, respectively.

The optical imaging performance of CVCEs and CFCEs is verified in the experiment. An emitting diode (LED) is used to illuminate the hollow letters "U" and "H", and the clear images can be observed on the image plane of the two compound eye structures, as shown in Fig. 4(a). As can be seen from Figs. 4(b)-4(e), in the central area, both compound eye structures can image clearly, but at the edge of FOV, there is a large angle between the optical axis of microlens unit and the incident light, so it does not have good imaging capability. In addition, because the optical axis of the microlens unit is not perpendicular to the CCD sensor plane, the image is distorted, but this distortion can be corrected by imaging algorithm. The central imaging was set as level 0, and we counted from the central axis outward and marked on Figs. 4(b) and 4(d). Compared with CVCEs (Fig. 4(b)), it can be found that CFCEs (Fig. 4(d)) has more identifiable imaging units than CVCEs, which further confirms that the special confocal design can improve the imaging quality of the edge of FOV and obtain more useful information. Meanwhile, it can be seen from Fig. 4 that CFCEs has larger imaging area than CVCEs. By calculating the imaging unit size of Figs. 4(c) and 4(e), it can be concluded that under LED irradiation, the unit imaging area of CFCEs increases by 26.6% (red-light LED), 24.9% (green-light LED) and 26.0% (blue-light LED), respectively, compared with CVCEs, which is beneficial to reducing image overlap.

In conclusion, we fabricate two kinds of 100% fill factor compound eye structures using DLW, including CVCEs and specially designed CFCEs, where the focal length of each microlens unit is equal to the distance between the microlens unit and the image sensor. It is found that compared with CVCEs, CFCEs can improve the relative focusing efficiency by about 7%, enlarge the imaging area by about 25%, and have better imaging quality at the edge of FOV. In addition, DLW can directly fabricate the compound eye structure on the surface of the device without subsequent transfer. Our research may have potential application to the optical detector, wide-FOV imaging system, integrated photonic chip and so on.

**Funding.** This work is supported by the Natural Science Foundation of China (No.62075073, 62135004 and 62075129), the Fundamental Research Funds for the Central Universities (No. 2019kfyXKJC038), State Key Laboratory of Advanced Optical Communication Systems and Networks, Shanghai Jiao Tong University (No. 2021GZKF007), and Key R & D project of Hubei Province (No. 2021BAA003).

**Acknowledgment.** The authors would like to thank Dr. Xing Feng and Dr. Tie Hu for the revisions and the suggestions on this letter.

**Disclosures**. The authors declare no conflicts of interest.